\documentclass[aip, jap, reprint]{revtex4-2}
\usepackage[tbtags]{amsmath}
\usepackage{amssymb}
\usepackage{amsfonts}
\usepackage[squaren]{SIunits}
\usepackage{braket}
\usepackage{slashed}
\usepackage{overpic}
\usepackage{subcaption}
\usepackage{booktabs}
\usepackage{multirow}
\usepackage{dcolumn}
\usepackage[utf8]{inputenc}
\usepackage[T1]{fontenc}
\usepackage{mathptmx}
\usepackage{etoolbox} 
\usepackage{duckuments} 
\usepackage{graphicx}
\usepackage{lineno}

\makeatletter
\def\@email#1#2{%
 \endgroup
 \patchcmd{\titleblock@produce}
  {\frontmatter@RRAPformat}
  {\frontmatter@RRAPformat{\produce@RRAP{*#1\href{mailto:#2}{#2}}}\frontmatter@RRAPformat}
  {}{}
}%
\makeatother

\newcommand{\md}{\mathrm{d}}
\newcommand{\me}{\mathrm{e}}

\begin{document}

\title{Analysis of Light Attenuation Length Measurement of a\\ High Quality Linear Alkylbenzene for the JUNO Experiment}

\author{Guojun~Yu}%
 \email{guojunyu@smail.nju.edu.cn}
 \affiliation{ 
    National Laboratory of Solid State Microstructure, Nanjing University%
}%
 \affiliation{ 
    School of Physics, Nanjing University, Hankou Road 22, 210093, Nanjing, China%
}%
\author{Jialiang~Zhang}%
\affiliation{ 
    National Laboratory of Solid State Microstructure, Nanjing University%
}%
 \affiliation{ 
    School of Physics, Nanjing University, Hankou Road 22, 210093, Nanjing, China%
}%
\author{Shuo~Li}%
\affiliation{ 
    National Laboratory of Solid State Microstructure, Nanjing University%
}%
 \affiliation{ 
    School of Physics, Nanjing University, Hankou Road 22, 210093, Nanjing, China%
}%
\author{Zifeng~Xu}%
 \affiliation{ 
    National Laboratory of Solid State Microstructure, Nanjing University%
}%
 \affiliation{ 
    School of Physics, Nanjing University, Hankou Road 22, 210093, Nanjing, China%
}%
\author{Lei~Zhang}%
\affiliation{
    School of Physics, Nanjing University, Hankou Road 22, 210093, Nanjing, China%
}%

\author{Aizhong~Huang}%
\affiliation{ 
    Jinling Petrochemical Corporation Ltd, 210033, Nanjing, China%
}%

\author{Ming~Qi}%
 \email{qming@nju.edu.cn}
 \affiliation{ 
    National Laboratory of Solid State Microstructure, Nanjing University%
}%
 \affiliation{ 
    School of Physics, Nanjing University, Hankou Road 22, 210093, Nanjing, China%
}%

\date{\today}

\begin{abstract}

Jiangmen Underground Neutrino Observatory (JUNO) is the next generation neutrino experiment which aims at neutrino mass hierarchy problem along with many other cutting-edge studies concerning neutrinos. Located \(700\metre\) underground in Jiangmen China, JUNO's central detector is an acrylic sphere filled with \(20\kilo\tonne\) liquid scintillator with linear alkylbenzene(LAB) as scintillator solvent. To ensure that an unprecedented energy resolution of \(\sigma_E/E \leqslant 3\%\) can be reached, LAB used in JUNO must have excellent transparency at the wavelength ranging from \(350\nano\metre\) to \(450 \nano\metre\).

In the past decade much effort has been devoted to the development of high transparency LAB based on the measurement of light attenuation length. Through a close cooperation with Jingling Petrochemical Corporation in  Nanjing, transparency of LAB samples prepared for JUNO has been improved progressively. However, this improvement is also pushing our apparatus towards approaching its measuring limit, undermining the credibility of our measurement. In order to get a result accurate and precise, an apparatus upgrading and a more detailed error analysis is inevitable. 

In this article, we present an analysis of how apparatus upgrading helps with decreasing measuring errors, and we conducted measurements using the new apparatus on several samples. A detailed error analysis is followed to validate the results. We propose to apply statistical methods featuring Monte Carlo simulation to estimate systematic uncertainties. Deviations caused by fit models is also considered and the overall uncertainty is obtained by combining two independent measurements. We finally report the light attenuation length of a newly improved LAB sample to be \(29.90 \pm 0.95 \metre\), which gives a new high of all the preceding samples we tested. This study may provide a strong evidence of JUNO's feasibility to reach its energy resolution.

\end{abstract}

\maketitle

\section{Introduction}

Following the Daya Bay's success of measuring neutrino oscillation parameter \(\theta_{13}\),\cite{DayaBay2012,DayaBay2014-1,DayaBay2014-2,DayaBay2015,DayaBay2016} Jiangmen Underground Neutrino Observatory (JUNO) is designed to tackle the neutrino mass hierarchy problem and many other exciting frontiers of neutrino research.\cite{JUNO-physics,JUNO2008,JUNO2009,JUNO2013} Located \(700\metre\) underground in Jiangmen with two nuclear power plants in Yangjiang and Taishan both \(53 \kilo\metre\) away, JUNO has a baseline that is optimal for mass hierarchy experiments.\cite{JUNO-physics,JUNO2008} JUNO's central detector is an arcrylic sphere with a diameter of \(35.4\metre\) immersed in a cylinder water tank. Roughly \(18,000\) microchannel plate photomultiplier tubes(PMT) and more than \(20,000\) other PMTs are equipped surrounding the sphere to detect and magnify weak \(\gamma\) signals produced via inverse beta decay \(\bar{\nu}_e + p \rightarrow e^+ + n\) process.\cite{JUNO-physics,JUNO2008} To be able to manifest the extremely small difference of different mass hierarchy models, JUNO's energy resolution should at least be at a level of \(\sigma _E/E \leqslant 3\%\) at \(1 \mega\electronvolt\), which amounts to a photon yield of \(1,100/\mega\electronvolt\).

As a consequence, optical transparency of liquid scintillator(LS) plays a key role in reaching such a high standard. JUNO uses an organic liquid scintillator with linear alkylbenzene(LAB) as solvent, PPO(\textit{2,5-diphenyloxazole}) as fluor and \textit{bis-}MSB(\textit{1,4-bis(2-methylstyryl)-benzene}) as wavelength shifter. LAB as solvent constitutes the majority of scintillator's mass so it dominates the light attenuation performance. LAB is an ideal organic target material. It is a hydrocarbon-rich compound with a high quantum yield, high flash point and many other outstanding features such as low toxicity, biological degradability, and relatively low price. Unfortunately, industrially produced LAB usually contains various impurities which greatly weaken its light penetrability. Thus, it cannot be emphasized how crucial it is to develop an LAB sample with a high transparency quality.

Numerous experiments as well as theoretical studies have been put into investigating light attenuation in fluid. Some focus on directly measuring light attenuation length,\cite{Huang2010, Goett2011, Cao2019, Yin2020, Yu2022} which is a comprehensive manifestation of absorption, scattering and even re-emission.\cite{Bohren1998, Huang2011, Xiao2010} Some focus on scattering process in specific, as a consequence of the fact that absorption and re-emission is weak in LAB and it is Rayleigh scattering that dominates the process.\cite{Zhou2015-1, Zhou2015-2} Thus it would suffice to measure Rayleigh scattering length and use it as an indication of material's transparency.\cite{Zhou2015-1, Zhou2015-2, Wurm2010, Liu2015} Note that since Rayleigh scattering is an elastic process it simply turns photons into different directions, making no contributions to decrease of transparency. Thus a lower scattering length is preferred.\cite{Zhou2015-1} The experiment presented in this article belongs to the former kind.

Through a close cooperation of more than 10 years with Jingling Petrochemical Corporation in Nanjing, LAB samples has been improved to show an increase of attenuatioan length from \(14\metre\) to \(25\metre\).\cite{Huang2010, Cao2019} Going further, in this article we report a result of \(29.90 \pm 0.95 \metre\) attenuation length of a newly improved LAB sample obtained after an apparatus upgrade. A detailed error analysis featuring Monte Carlo method is also given to validate this result.

\section{Experiment}

\subsection{Attenuation length}

Light attenuation length \(L\) of some substance is defined as the longest length of a beam can travel in this substance until its intensity drops to \(1/\me\) of the incident. The basic method of measuring the attenuation length of a liquid substance is fitting intensity data collected at different liquid levels to the Beer--Lambert law:\cite{Beer1852}

\begin{equation}\label{eqn:lambert}
    I = I_0\me^{-\frac{x}{L}}
\end{equation}
where \(I_0\) is the incident intensity, \(x\) is the path beam travels. 

\begin{figure}[htp!]
    \centering
    \includegraphics[scale = 0.3]{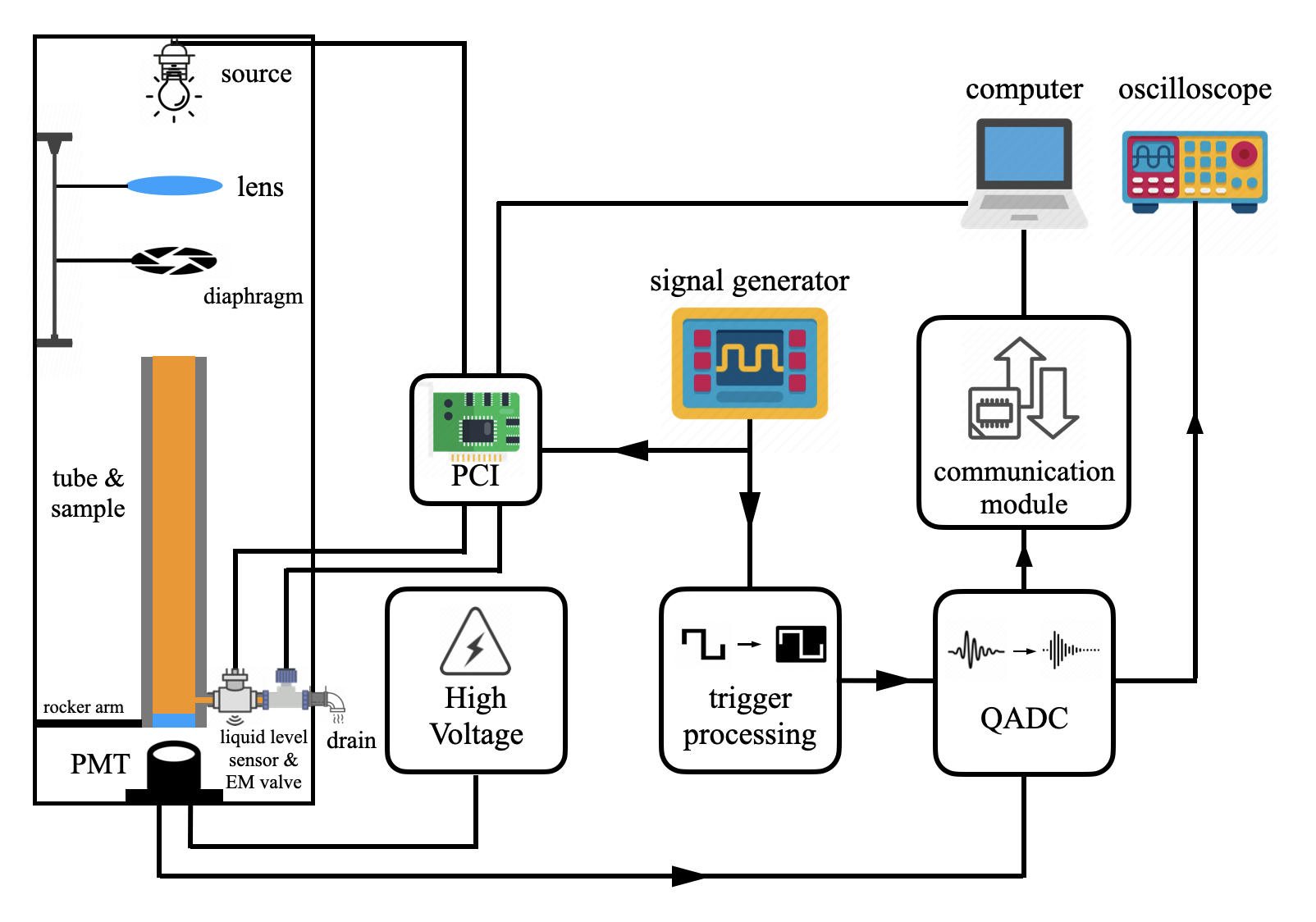}
    \caption{The whole experiment apparatus is placed in a dark room with electromagnetism screening to make sure its stability}
    \label{fig:apparatus}
\end{figure}

Obviously if a set of liquid levels and intensity data is provided, \(L\) can be obtained by fitting. Designed based on this, our apparatus comprises two major parts: optical measuring system and data acquisition system(DAQ). As is shown in figure~\ref{fig:apparatus}, the optical measuring system is all contained in a dark closet. At the top of the closet an LED source with a \(430\nano\metre\) peak wavelength is installed. Aligned below is an adjustable lens that parallelizes beam, a plano-convex spherical lens with focus \(0.15\metre\), a diaphragm, a one-meter-tube used as a sample container, and a PMT(CR135 from Hamamatsu) that detects light intensity and outputs electronic signals for later processing. Note that attenuation length varies with incident light with different wavelengths so that LED source must have an appropriate wavelength range. In JUNO experiment, photons arriving at PMTs will be at a wavelength of roughly \(430\nano\metre\) so this is set to be the LED's peak wavelength. At the bottom side of the tube there is a conduit in which a pressure transmitter and an electromagnetic valve are plugged. By controlling the valve, sample in the tube can be set to different liquid levels. The intensity collected by PMT and the level data provided by the transmitter form a dataset to be fit to finally get the attenuation length.

However, considering the randomness in evaluation, a single measurement of intensity is far from enough. Through a PCI board, the LED source is controlled to pulse with an \(800\hertz\) on-off signal generated by a signal generator so that in one minute a dataset of \(48,000\) data points are generated. A trigger signal at the same frequency is also generated to accommodate the DAQ. The original trigger will be processed and sent as a square wave to a charge-to-amplitude converter(V965 from CAEN). Finally, the analog intensity signal will be transformed to a \(12\) bit digital ADC channel and then saved. Eventually around \(19,000\) data can be recorded in one minute due to the memory limit and the intensity is determined to be the mean of these \(19,000\) data through a Gaussian fit.

Inasmuch as the analog signal is processed by an ADC, it is not the absolute intensity but a digital ADC channel number ranging from \(0\) to \(2^{12}\) that is obtained. However, since the value of ADC channel is proved to be linearly dependent on intensity with an intercept resulted from electronical pedestal(abbreviated as ped below), fitting ADC channel leads to no difference in the final result than that obtained by fitting absolute intensity directly. Thus, for the balance of this article \(I\) will be used to denote ADC channel but still be referred to as intensity just for later convenience.

\subsection{Preparations}

Before the experiment, a full check on the equipment is always indispensable. It must be ensured that one-meter-tube is not contaminated, optical system is tuned to its best state, and every electronical device works properly. Apparatus calibration is done by performing a real measuring using branded purified water as sample. Taking advantage of the stability of purified water production, deviations of apparatus' performance can be detected.

\begin{figure*}[htp!]
\centering
    \begin{subfigure}{\textwidth}
        \includegraphics[width = \textwidth]{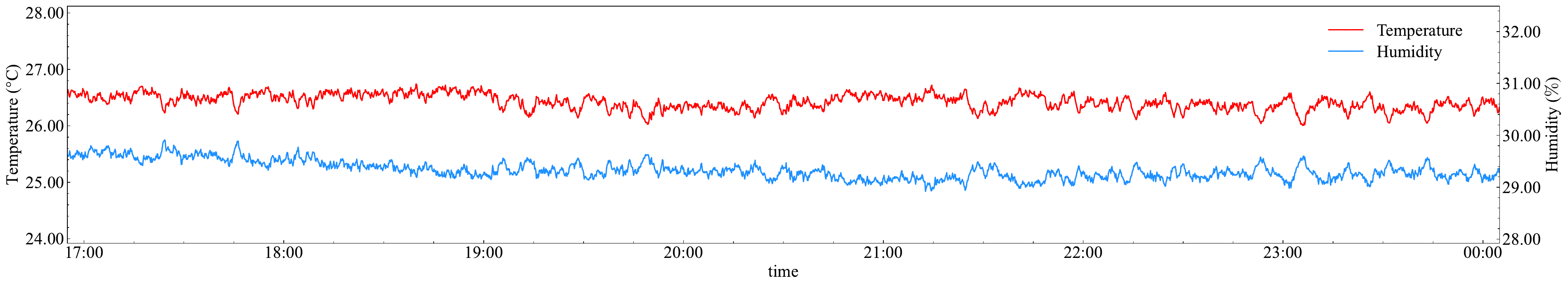}
        \caption{Temperature and relative humidity in the lab from 17:00 to 24:00. The maximal difference of temperature and that of humidity are respectively \(0.8\celsius\) and \(1\%\).}
        \label{}
    \end{subfigure}
    \scalebox{0.965}[1]{%
    \begin{subfigure}{\textwidth}
        \includegraphics[width = \textwidth]{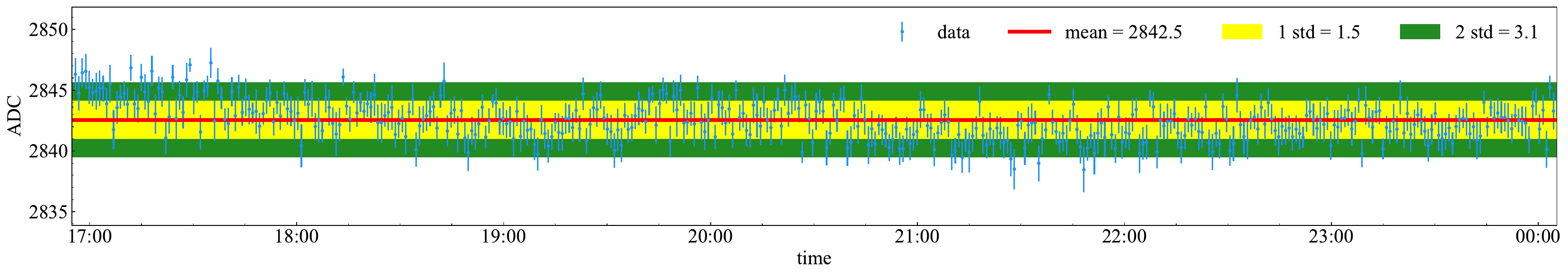}
        \caption{Intensity in the same time period. Fluctuation is relatively small when temperature and humidity is relatively stable.}
        \label{}
    \end{subfigure}}\hspace{6mm}
\caption{Environment's influence on intensity.}
\label{fig:thermohygro stable}
\end{figure*}

The stability of lab environment is important as well. Figure \ref{fig:thermohygro stable} shows a \(7\hour\) surveillance of lab temperature and humidity with AC turned on using a thermohygrometer, which has a resolution of \(0.01\celsius\) and \(0.024\%\)RH. It can be seen that under such circumstances the fluctuation is small. Though it is impossible to keep temperature and humidity a constant, it is always possible to perform experiment in a relatively short time so that the fluctuations of the environment is of little importance.

To guarantee the reliability of the experiment, it would only start after all the preparations are done. In this article 4 batches of newly developed LAB samples tagged as NJ64, NJ66, NJ67, and NJ68 are tested. Besides, a previously tested batch NJ44 is also used as a cross-check.

\section{Apparatus Upgrading}

\subsection{Necessity of upgrading}

For JUNO the longer attenuation length observed in some new sample the better. However, this increasing of attenuation length also creates a challenge for the measurement and pushing our apparatus towards its measuring limit, affecting both accuracy and precision of the result. The apparatus is claimed to reach its measuring limit in the following sense: collect a dataset and graph it, if there are two neighboring points coming too close to each other that their error bars overlap in the ordinate, it is believed that statistically they cannot be clearly distinguished from each other since this introduces ambiguity and makes the fitting result unreliable. To show this quantitatively, suppose \(n\) data points \(\{(x_i, I_i)\}\) are collected in a measurement, we define the relative intensity resolution \(\Delta I/I_0\) of a fit model \(I = f(x; I_0)\) as the minimal relative difference \(\Delta I/I_0 = \displaystyle\min_{|i - j| = 1}|I_i - I_j|\) of two neighboring points \((x_i, I_i)\) and \((x_j, I_j)\). In our case the fit model is the Beer--Lambert law \eqref{eqn:lambert}, and since \(\Delta x = |x_i - x_j|\) is two orders of magnitude smaller than \(L\), it is natural to write

\begin{equation}
    \frac{\Delta I}{I_0} = \left|\me^{-\frac{x_i}{L}} - \me^{-\frac{x_j}{L}}\right| \approx \frac{\Delta x}{L}
\end{equation}
In order to ensure that each data point is statistically distinguishable from the others, it is required that
\begin{equation}
    \Delta I \geqslant 2\varepsilon_{I}
\end{equation}
where \(\varepsilon_{I}\) is the uncertainty of intensity. This gives the maximal attenuation length the apparatus can give with accuracy and precision to some extent
\begin{equation}\label{eqn:resolution}
   L_{\max} \approx \frac{\max I_0}{2\varepsilon_I}\Delta x
\end{equation}
Note that this limit can always be avoided by simply extending the tube. So it has to be pointed out that this analysis only works when an extension is impossible and reduction of the number of data points is unacceptable, as is our case.

Usually intensity data will be collected at \(10\) different levels uniformly distributed in a distance of \(1\metre\) so \(\Delta x = 0.1\metre\). In order to get the uncertainty \(\varepsilon_I\) and the maximal \(I_{\max} = \max\, I_0\) in theory, a stability test involving large number of repeated measurements at some specific liquid level has been conducted. Histograms in figure \ref{fig:res-before-hist} and \ref{fig:sigma-before-hist} show the distribution of estimated means and standard deviations of roughly 200 repeating measurements at the same liquid level. The standard deviation of the mean values of intensity speaks for \(\varepsilon_I\), and the mean of the standard deviations of intensity gives information about \(I_{\max}\). Fit results show that \(I_{\max} = 2,900\) and \(\varepsilon_I = 4.5\). Note that \(\varepsilon_I\) provided here considers only the fluctuations of intensity itself. In reality \(\varepsilon_I\) would be slightly larger because of other errors such as error from fitting. Here a conservative estimation setting \(\varepsilon_I = 5\) is applied. With all parameters provided, the extreme attenuation length under the present circumstance is roughly

\begin{equation}
   L_{\max} \approx 29 \metre
\end{equation}

From equation \eqref{eqn:resolution} it can be easily seen that \(L_{\max}\) increases when \(I_0\), \(\Delta x\) increases or \(\varepsilon_I\) decreases. Since the total length of the tube cannot be altered, increasing \(\Delta x\) is tantamount to lowering the number of data point \(n\), which would drastically undermines goodness-of-fit. Thus the only two choices left are raising \(I_0\) and reducing \(\varepsilon_I\). Next we investigate two upgrades that both increase \(I_0\) and decrease \(\varepsilon_I\) at the same time. With these two upgrades not only would the maximal attenuation length \(L_{\max}\) this apparatus can measure be lifted up but also the uncertainties would be lowered.

\subsection{Collimatoed fibre-coupled source}

In the original apparatus the light source is simply an LED mounted at the top of the closet. The half divergence angle is up to \(6\degree\) and a diaphragm is needed to confine and block stray light. Considering this along with other causes such as unavoidable aberrations resulted from spherical lens, it is nearly impossible for light beam to propagate from the LED source to PMT in a straight line. This undermines the validity of fit using the ideal Beer--Lambert law, and at the same time enhances the fluctuations of intensity and brings down the \(I_0\). Thus, we come up with a new collimated fibre-coupled LED source design that aims at concentrating light and keep it non-divergent as much as possible.
\begin{figure*}[htp!]
    \centering
        \begin{subfigure}{0.4\textwidth}
            \includegraphics[width = \textwidth]{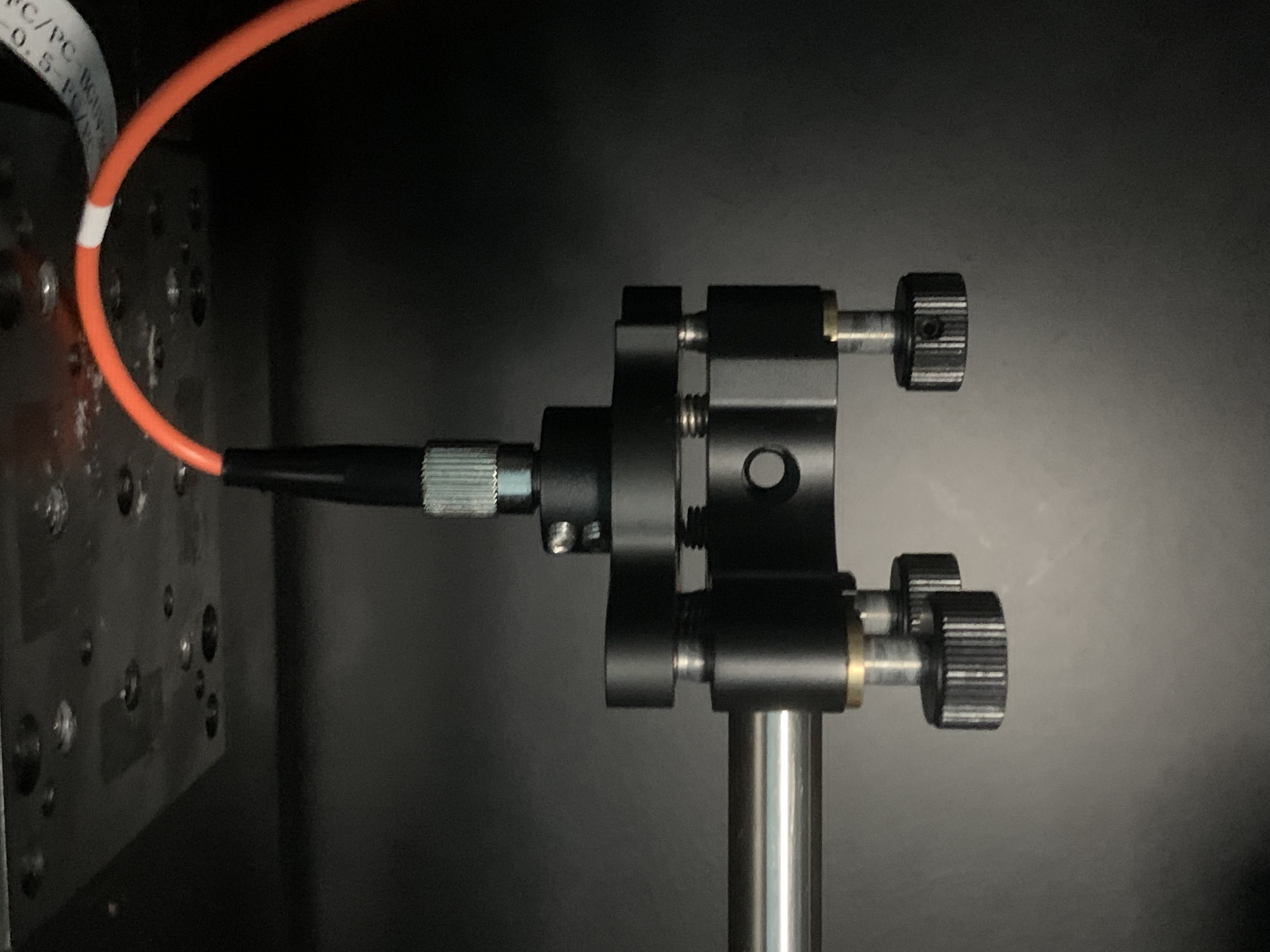}
            \caption{Fiber held by a mount which can be conveniently adjusted by the screws, helping to align the optical path.}
            \label{fig:source}
        \end{subfigure}
        \begin{subfigure}{0.46\textwidth}
            \includegraphics[width = \textwidth]{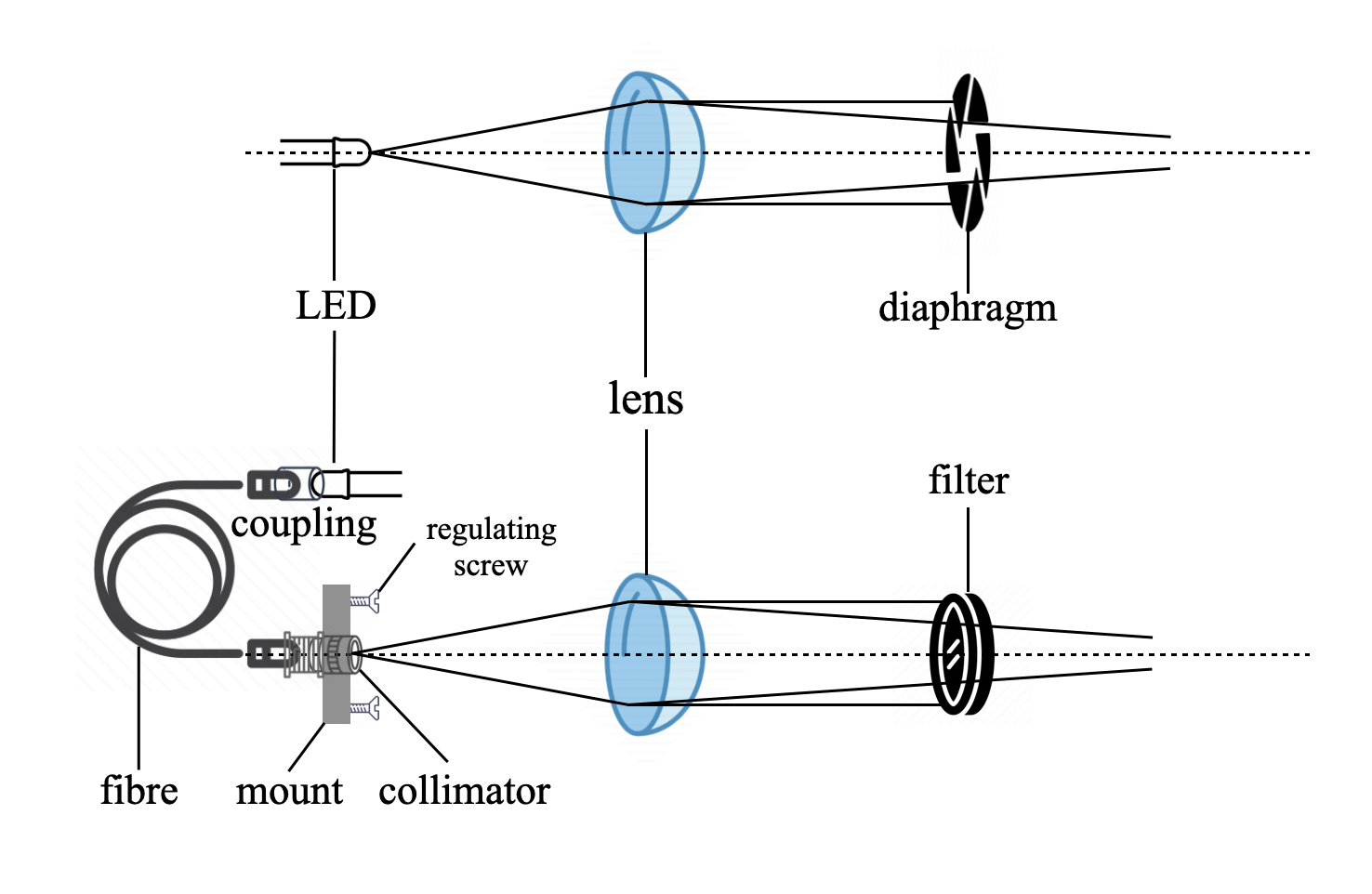}
            \caption{A comparison of two designs. The new collimated fibre source renders diaphragm unnecessary.}
            \label{fig:optical}
        \end{subfigure}
    \caption{The newly designed fibre-coupled LED source.}
    \label{fig:}
\end{figure*}

In the new apparatus, LED is coupled to an optical fibre. When it comes to curtailing divergence angle, usually the smaller the numerical aperture(NA) of a fibre the better, but a too small NA means a too small fibre diameter, leading to a relatively low light flux. Bound by this trade-off, we tested various optical fibres and decided to use a multi-mode fibre with an NA of \(0.22\) and a diameter of \(300\micro\metre\) for an optimal performance. A tailor-made collimator is attached to the other end of the fibre. The collimator is held by a three-dimensionally adjustable mount as is shown in figure \ref{fig:source}. A comparison diagram of the source before and after is drawn in \ref{fig:optical}.

\begin{figure*}[htp!]
    \centering
        \begin{subfigure}{0.3\textwidth}
            \includegraphics[width = \textwidth]{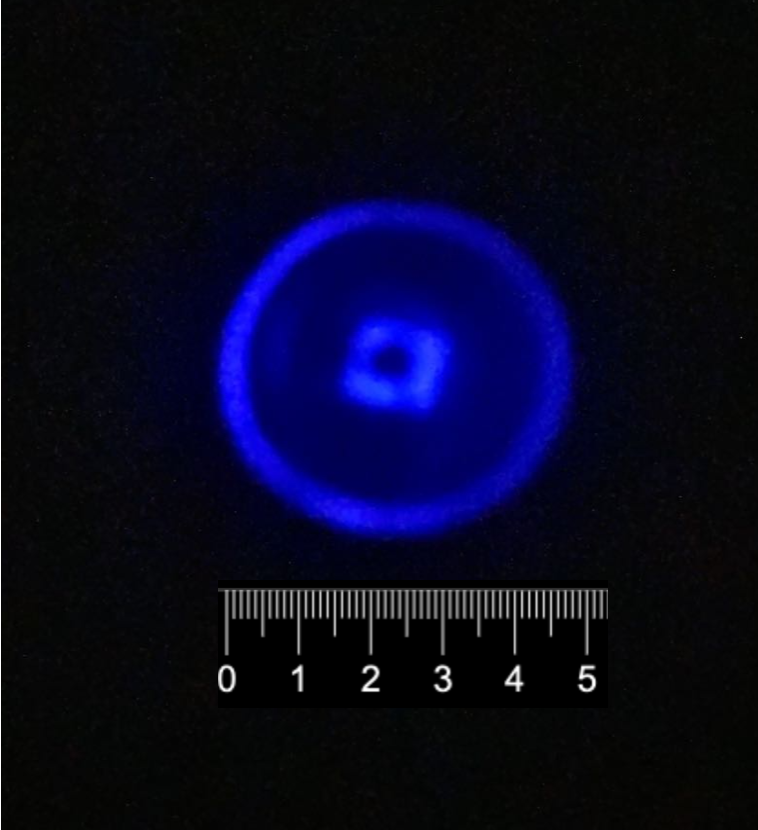}
            \caption{before}
            \label{fig:spot size before}
        \end{subfigure}
        \begin{subfigure}{0.3\textwidth}
            \includegraphics[width = \textwidth]{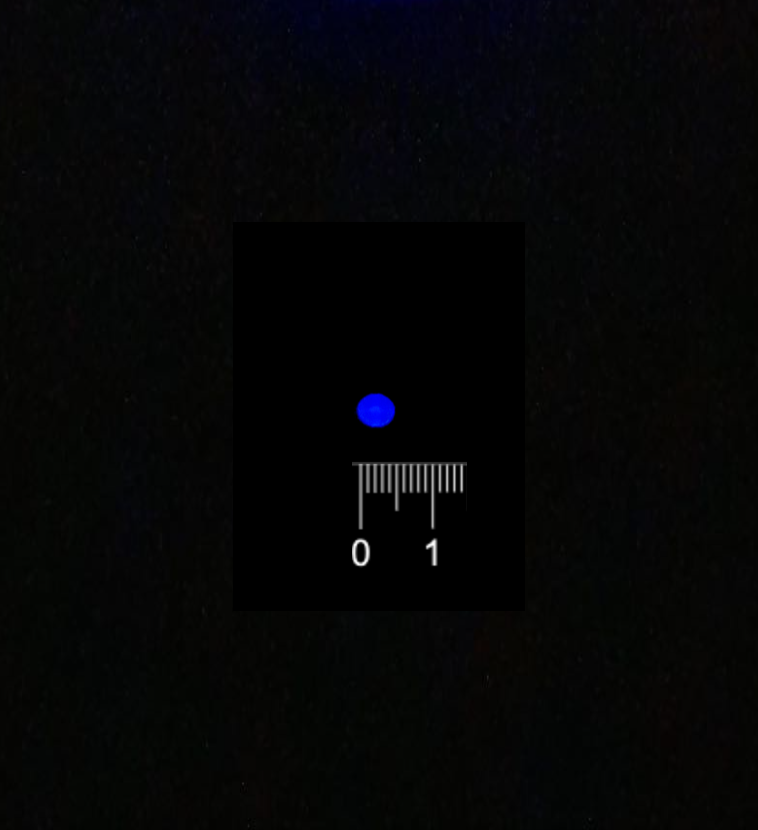}
            \caption{after}
            \label{fig:spot size after}
        \end{subfigure}
    \caption{Spot size(\centi\metre) before and after the upgrade. Note that in figure \ref{fig:spot size before} only the central image contributes to intensity value, the outer ring caused by diaphragm is excluded in real measurement.}
    \label{fig:spot size}
\end{figure*}

The direct effects of this optical upgrade is that the size of the spot on PMT's receiving window is shrunk tremendously, as is shown in figure \ref{fig:spot size}. A smaller light spot indicates a more parallel beam, hence less scattered light which gives more validation of applying Beer--Lambert law and save lots of effort for fine tuning optical system into an ideal condition. A smaller light spot also means more concentrated light energy, which allows a higher \(I_0\), and less perturbations, which reduces \(\varepsilon_I\).

\subsection{PMT upgrading}

An obvious reason that \(I_0\) cannot get higher lies in the fact that DAQ device has a limit in the maximal intensity it can measure. At each liquid level intensity \(I\) observes a normal distribution with a mean of \(\mu\) and standard deviation \(\sigma\), which are estimated from \(19,000\) ADC data. A fit with less than \(1\) in \(19,000\) data is lost in an estimation indicates that every data within the range of \(\mu \pm 4\sigma (P(|x - \mu| \leqslant 4\sigma) = 99.99266\%)\) must be included (\(1 - 99.99266\% \approx 1/15787\)). Thus to avoid any possible bias because of loss in data it is required that \(I_{\max} \leqslant 4096 - 4\sigma - \text{ped}\). For our apparatus \(\text{ped} = 273\). Thus we have \(I_{\max} = 3823 - 4\sigma\).

PMT plays a key role in collecting ADC channel data and its performance determines how well ADC data's distribution can be. A better PMT will have a faster time response, finer time resolution, and a more sensitive and uniform cathode, which means a higher robustness against different perturbations that causes ADC channels to show a wider spread. Naturally this would decrease the distribution's \(\sigma\) so it increases \(I_{\max}\). Moreover, a smaller \(\sigma\) also means a smaller \(\varepsilon_I\).

\begin{table*}[htbp!]
    \centering
    \caption{Key parameters of CR135 and R7724}
    \label{tab:PMT compare}
    \scalebox{1}[1]{%
    \begin{tabular}{ccccccccc}
        \toprule
        ~ & \multicolumn{2}{c}{spectral response(\(\nano\metre\))} & ~ & \multicolumn{2}{c}{supply voltage (\(\volt\))} & ~ &\multicolumn{2}{c}{luminous (\(\micro\ampere/\textrm{lm}\))}  \\ \cmidrule{2-3}\cmidrule{5-6}\cmidrule{8-9}
        type & range & maximum & rise time(\(\nano\second\)) & operational & maximal & dark current(\(\nano\ampere\)) & cathode & anode  \\ \midrule
        CR135 & 300\(\sim\)650 & 420 & 7 & 1,250 & 1,500 & 30 & 60 & 25\(\times 10^8\) \\
        R7724 & 300\(\sim\)650 & 420 & 2.1 & 1,750 & 2,000 & 6 & 90 & \(3\times 10^8\) \\ \bottomrule
    \end{tabular}}
\end{table*}

To achieve this we replace the old-fashioned CR135 from Hamamatsu to R7724 with a faster time response and lower noise. Both PMT's peak cathode sensitivity is arrived at \(430\nano\metre\). Other parameters are shown in Table \ref{tab:PMT compare}. A new voltage divider is also designed to accommodate the new PMT. This voltage divider applies an iterative voltage distribution of resistors with \(4\) decoupling capacitors in series so that it has the biggest gain and shows the best performance when conducted in counting mode. A total resistance of \(3.5\mega\ohm\) gives a current of \(500\micro\ampere\) operated under a voltage of \(1,750\volt\).

\begin{figure*}[htp!]
\centering
    \begin{subfigure}{0.4\textwidth}
        \includegraphics[width = \textwidth]{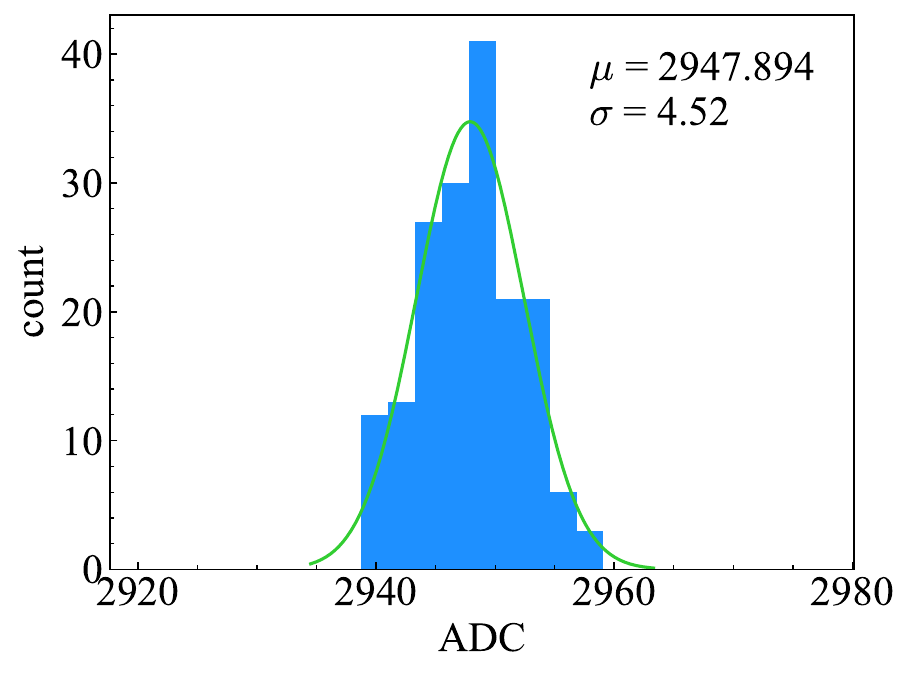}
        \caption{CR135 without fibre and collimator}
        \label{fig:res-before-hist}
    \end{subfigure}
    \scalebox{1}[0.99]{%
    \begin{subfigure}{0.4\textwidth}
        \includegraphics[width = \textwidth]{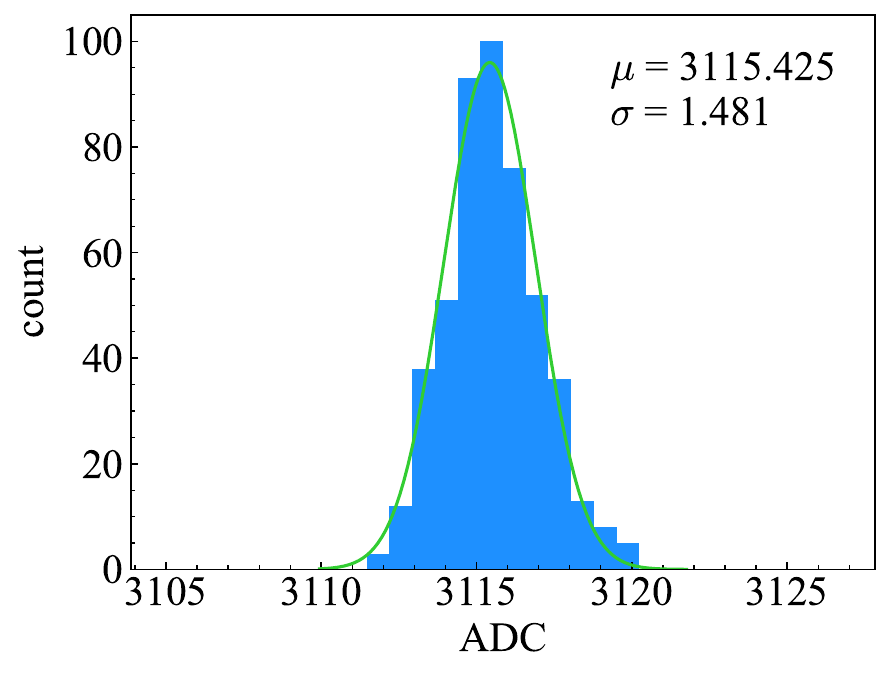}
        \caption{R7724 with fibre and collimator}
        \label{fig:res-after-hist}
    \end{subfigure}}
    \scalebox{0.965}[0.97]{%
    \begin{subfigure}{0.4\textwidth}
        \includegraphics[width = \textwidth]{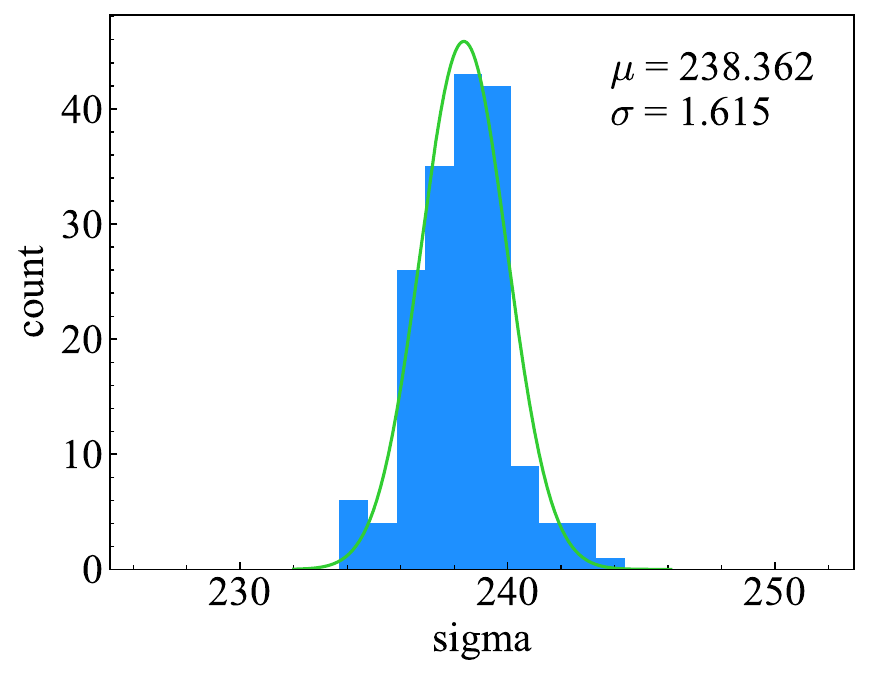}
        \caption{CR135 without fibre and collimator}
        \label{fig:sigma-before-hist}
    \end{subfigure}}\hspace{4mm}
    \scalebox{1}[0.99]{%
    \begin{subfigure}{0.4\textwidth}
        \includegraphics[width = \textwidth]{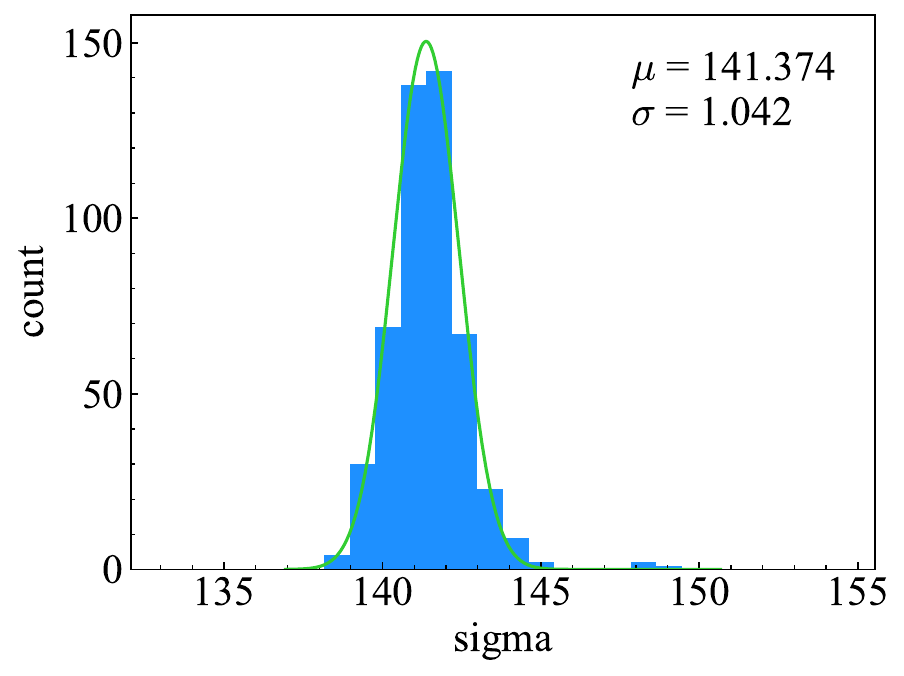}
        \caption{R7724 with fibre and collimator}
        \label{fig:sigma-after-hist}
    \end{subfigure}}
\caption{Distributions of mean values(upper) and standard deviations(lower) provided by Guassian fit of intensities. \ref{fig:res-before-hist} and \ref{fig:sigma-before-hist} are tested using CR135 and no fibre and collimator. \ref{fig:res-after-hist} and \ref{fig:sigma-after-hist} are tested using R7724 with the new collimated fibre-coupled source.}
\label{fig:resolution compare}
\end{figure*}

Similarly, a stability test has been performed with every parameter fixed except for PMT and light source. Histogram in figure \ref{fig:res-after-hist} clearly shows a smaller variance of intensity than in figure \ref{fig:res-before-hist}. Fit results present that \(\varepsilon_I \approx 1.5\) and \(\sigma \approx 141\), which means \(I_{\max} = 3,200\). Again a more conservative estimation is adopted by setting \(\varepsilon_I = 2\). with these and \(\Delta x = 0.1\metre\) all inserted into equation \eqref{eqn:resolution}, the limit attenuation length under the present circumstance is
\begin{equation}
   L_{\max} \approx 80\metre
\end{equation}

After upgrading \(L_{\max}\) is lengthened high enough that little influence is exerted on the accuracy of the measurement. The following section would show how these upgrades also improve the precision of the measurement.

\section{Results and Error Analysis}

\begin{figure*}[htp!]
    \centering
        \begin{subfigure}{0.47\textwidth}
            \includegraphics[width = \textwidth]{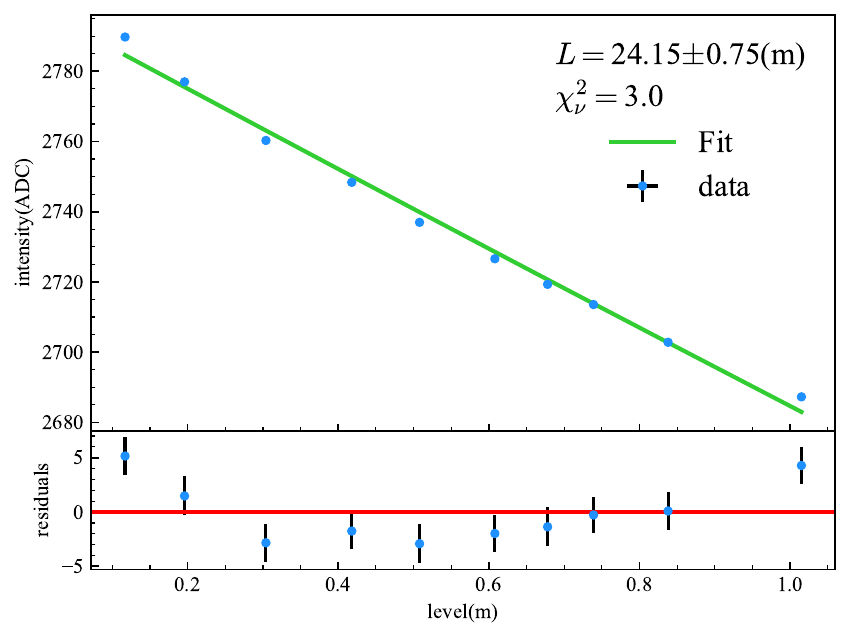}
            \caption{NJ64}
            \label{NJ64 before}
        \end{subfigure}
        \begin{subfigure}{0.47\textwidth}
            \includegraphics[width = \textwidth]{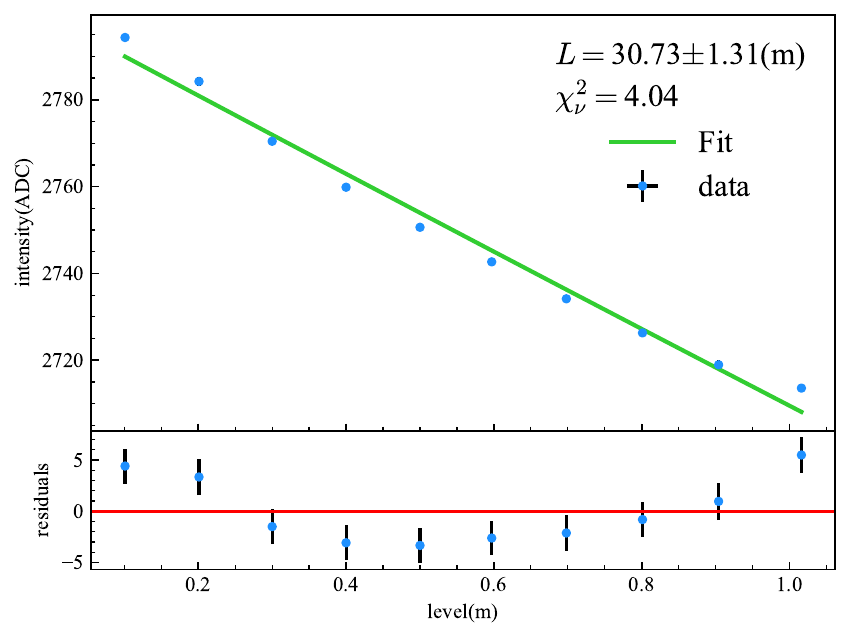}
            \caption{NJ66 test 1}
            \label{NJ64-1 after}
        \end{subfigure}
        \begin{subfigure}{0.47\textwidth}
            \includegraphics[width = \textwidth]{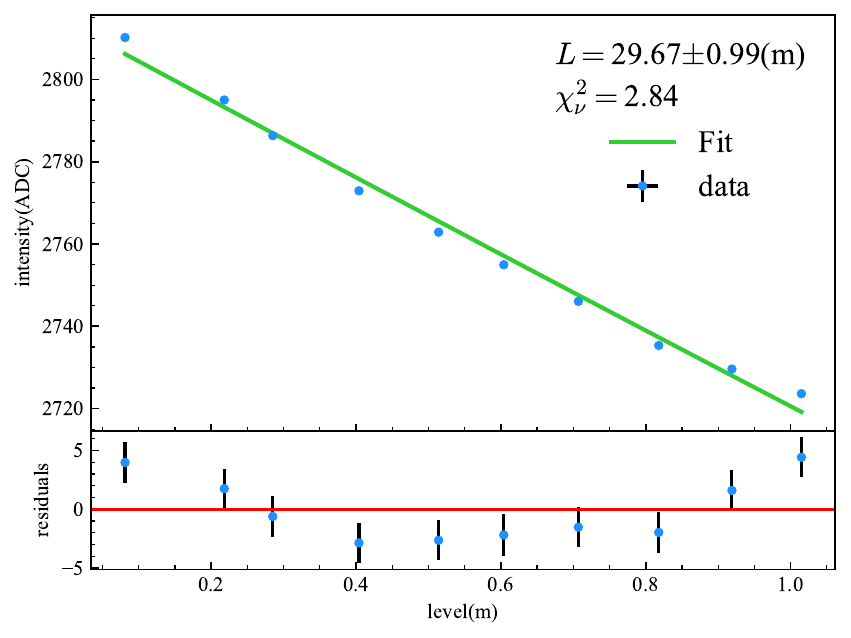}
            \caption{NJ66 test 2}
            \label{NJ64-2 after}
        \end{subfigure}
        \begin{subfigure}{0.47\textwidth}
            \includegraphics[width = \textwidth]{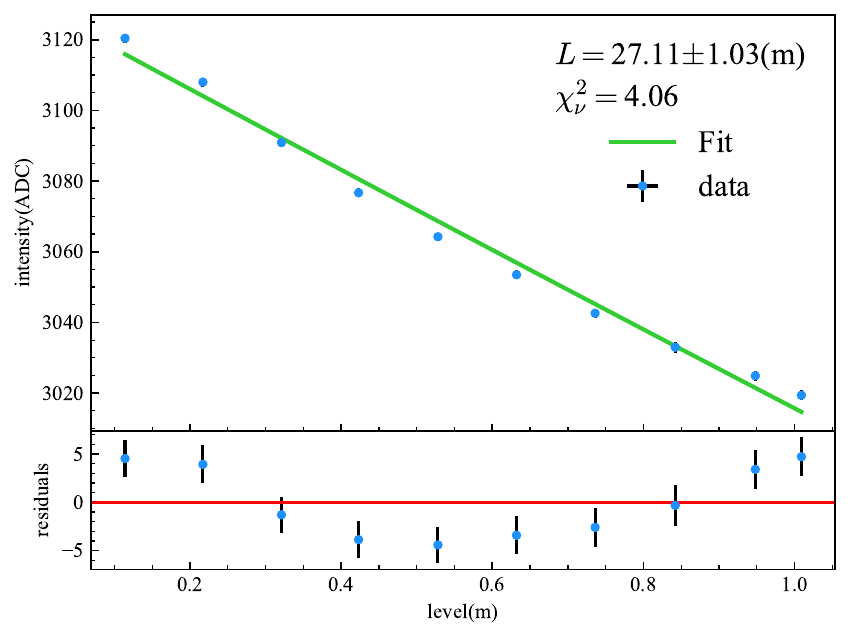}
            \caption{NJ67}
            \label{NJ66 before}
        \end{subfigure}
        \begin{subfigure}{0.47\textwidth}
            \includegraphics[width = \textwidth]{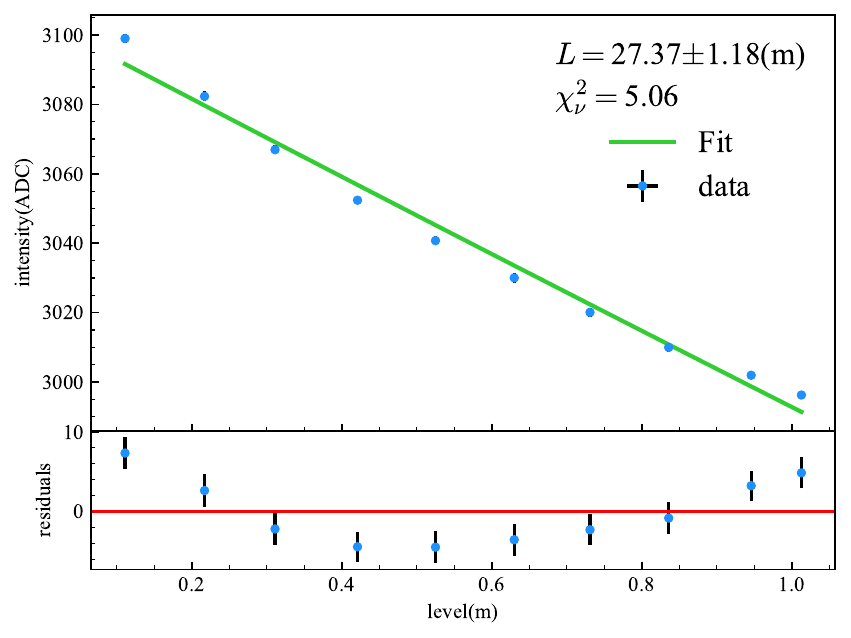}
            \caption{NJ68}
            \label{NJ66 after}
        \end{subfigure}
        \begin{subfigure}{0.47\textwidth}
            \includegraphics[width = \textwidth]{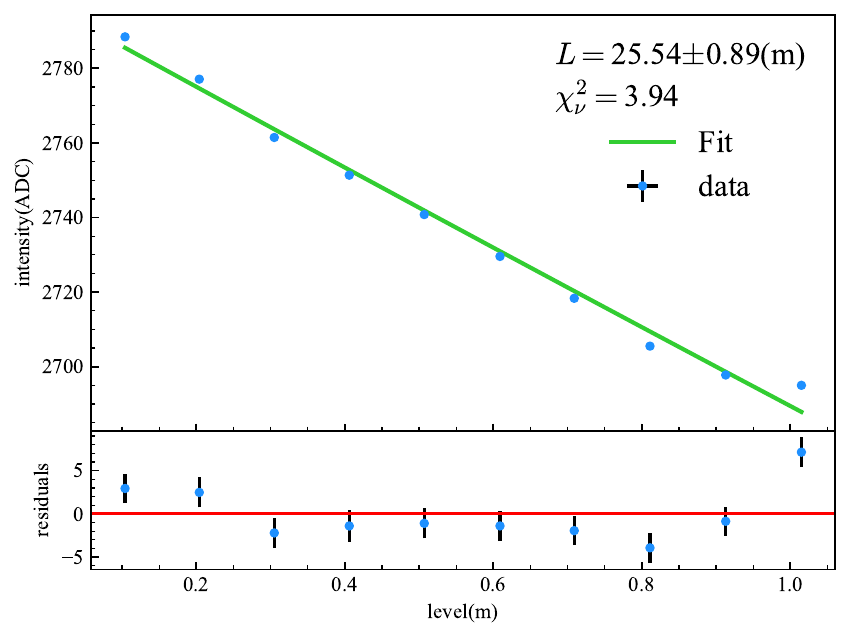}
            \caption{NJ44}
            \label{NJ44 after}
        \end{subfigure}
    \caption{Fit results of LAB samples. Reduced chi-square \(\chi_{\nu}^2\) is chi-square over degrees of freedom. Note that in fit plots error bars are too small to be seen but can be seen in residual plots.}
    \label{fig:results}
\end{figure*}

With the aid of upgraded apparatus, 4 batches of newly developed LAB samples, namely NJ64, NJ66, NJ67 and NJ68 have been measured. Each sample differs from each other in respect of impurities. To prove the robustness of the experiment, an old batch of NJ44 has been re-measured as well. Figure \ref{fig:results} summarizes results for each sample.

Compared to the best results we get before,\cite{Cao2019} it is implied in the figures that every new sample evidently shows a better transparency, especially NJ66, whose attenuation length in the test 1 exceeds \(30\metre\) for the first time. Re-measurement using the former NJ44 sample seems to further mutually confirms the reliability of the results. Meanwhile, the residuals appear to show a pattern of high ends and low abdomen. To figure out how this phenomenon affects the result a physical analysis of the fit model along with a detailed error analysis is also given.

Researches of attenuation length measurement usually report errors of the Beer--Lambert fit as the final result without further dissecting it into systematic and statistical errors. Errors from the abscissa are rarely considered either. To tackle these problems, the error analysis in this article pays more attention to statistical methods such as Monte Carlo simulation, rendering the results more informative.

The uncertainty of attenuation length \(\varepsilon_L\) in our experiment comprises three parts: systematic uncertainty \(\varepsilon_L^{\text{sys}}\), statistical uncertainty \(\varepsilon_L^{\text{sta}}\), and model uncertainty \(\varepsilon_L^{\text{model}}\). Systematic and statistical uncertainty will both be reflected in the fit error if systematic errors of intensity and liquid level are contained in the error bars when fitting. In the following section uncertainties of intensity and liquid level will be examined separately in detail first and then combined in the Monte Carlo simulation to give the systematic uncertainty. Model uncertainty is introduced to account for the residual pattern and will be discussed later.

\subsection{Uncertainties of intensity}

Obtained by fitting \(19,000\) data to the Gaussian model, the uncertainty of intensity \(\varepsilon_I\) has two sources: statistical \(\varepsilon_I^{\text{sta}}\) and systematic \(\varepsilon_I^{\text{sys}}\). \(\varepsilon_I^{\text{sta}}\) is given by Gaussian fitting straight away while \(\varepsilon_I^{\text{sys}}\) has to be determined by experiment since it results from nuisance parameters reflecting effects of some inevitable interferences, such as fluctuation of LED's emission power, oscillation of building, stray light background, and performance of electronical device, etc. These factors can all be taken into consideration by collecting intensity data at some fixed level \(x_i\) in a fairly long testing cycle. In figure \ref{fig:resolution compare}, the two upper plots present distributions of mean and standard deviation of intensities in NJ66 sample at some liquid level. It is clearly manifested in the figures that after the upgrades that \(\varepsilon_I^{\text{sys}}\) drops drastically, which means precision of \(I_i\) gets better. Note that in principle this measurement should be done for every single sample at every liquid level. Considering the obvious inconvenience and bother this test has only been done once for each of the two representative samples NJ66 and purified water. The test confirms that \(\varepsilon_I^{\text{sys}}\) varies little with liquid levels or samples and is a good indication of systematic influence. The uncertainty of intensity at level \(x_i\) is denoted as

\begin{equation}
    \varepsilon_{I}(x_i) = \sqrt{\varepsilon_I^{\text{sys}}(x_i)^2 + \varepsilon_I^{\text{sta}}(x_i)^2}
\end{equation} 

\subsection{Uncertainties of liquid levels}

Liquid levels are calculated from voltage data provided by a pressure transmitter. The transmitter responses with hydrostatic pressure \(p\), which is given by the Bernoulli's equation
\begin{equation}
    \frac{p}{\gamma} + \frac{v^2}{2g} + z = 0
\end{equation}
where \(z\) is depth and \(x = -z + h_0\) with \(h_0\) the extra length below the conduit and \(\gamma = \rho g\) with \(\rho\) the mass density of liquid and \(g\) gravity acceleration. For the hydrostatic case here liquid velocity \(v = 0\). 

\begin{figure}[htp!]
    \centering
    \includegraphics[scale = 0.6]{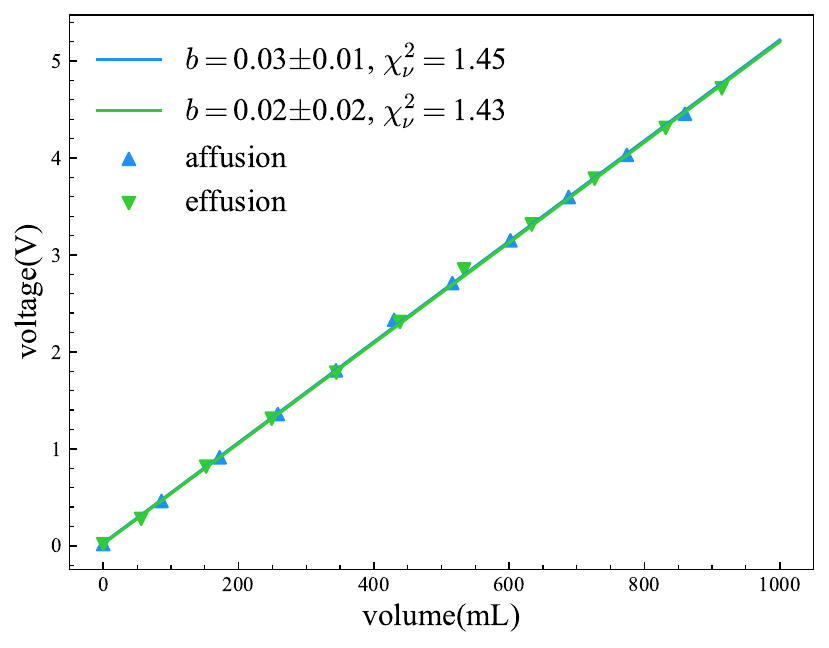}
    \caption{Linearity test of pressure transmitter. Two methods both give a \(\chi^2_{\nu} < 2\), thus the linearity is proved.}
    \label{fig:transmitter}
\end{figure}

Usually a qualified transmitter's response should be linear. If it is not, substantial errors would appear. We tested the linearity of the transmitter by adding and draining water in the tube successively and fit its responses to a straight line. Fit results in figure \ref{fig:transmitter} shows estimated intercepts of both tests are close to \(0\) and \(\chi^2_{\nu}\) close to \(1\), indicating that linearity of the transmitter is guaranteed. Thus the error of liquid level \(\varepsilon_x\) is simply

\begin{equation} 
    \varepsilon_x = \frac{1}{\gamma}\varepsilon_p 
\end{equation}

PCM300's data sheet reports a relative error \(0.5\%\)FS, so that \(\varepsilon_x = 0.5\%x\).

\subsection{Monte Carlo simulation}

In principle, if errors exist for both coordinates, the function to be minimized is generalized to be a weighted sum of residuals' squares of both variables.\cite{Deming} Under some circumstances this sum does not even have a chi-square distribution, introducing so much trouble in specifying confidence limits. Fortunately, taking advantage of the approximate linear behavior of the exponential model when the argument \(\Delta x /L\) is small, the chi-square can be written as\cite{Ifan}

\begin{equation} 
    \chi^2 = \sum_{i = 1}^n \frac{\left(I_i - f(x_i; L, I_0)\right)^2}{\varepsilon_{I_i}^2 + k_i^2\varepsilon_{x_i}^2}
\end{equation}
where \(k_i = -I_0/L\me^{-x_i/L}\) is the slope of the model at \(x_i\). Though simplified, this chi-square is still complicated to optimize due to the non-linear dependence of \(\chi^2\) on parameters \(L\). 

A Monte Carlo simulation is introduced to overcome this problem. In one measurement a dataset of several liquid levels, intensities and corresponding uncertainties \(\{(x_i, I_i, \varepsilon_I(x_i)\}\) is obtained and gives only one result of \(L\). In order to exploit the most of this dataset, we generate a new dataset \(\{(x_i, I'_i, \varepsilon_{I'}(x_i)\}\) from the original data by sampling from a set of Gaussian distributions \(I'_i \sim \mathcal{N}(I_i, \varepsilon_I'(x_i))\) for each \(x_i\). Here \(\varepsilon_{I'}(x_i) = \sqrt{\varepsilon_{I_i}^2 + k_i^2\varepsilon_{x_i}^2}\). The probability density function of \(I'_i\) is thus

\begin{equation} 
    p(y) = \frac{1}{\sqrt{2\pi}\varepsilon_{I'}(x_i)}\me^{-\frac{(y-I_i)^2}{2\varepsilon^2_{I'}(x_i)}}
\end{equation}

Fitting this generated dataset provides a new estimated \(L\). Repeat this process enough times so a substantial number of \(L\) is obtained. The central limit theorem guarantees that this distribution is almost Gaussian. Thus it is reasonable to fit this Gaussian distribution and claim the standard deviation as an estimation of the systematic uncertainty \(\varepsilon_L^{\text{sys}}\).

\begin{figure*}[htp!]
\centering
    \begin{subfigure}{0.45\textwidth}
        \includegraphics[width = \textwidth]{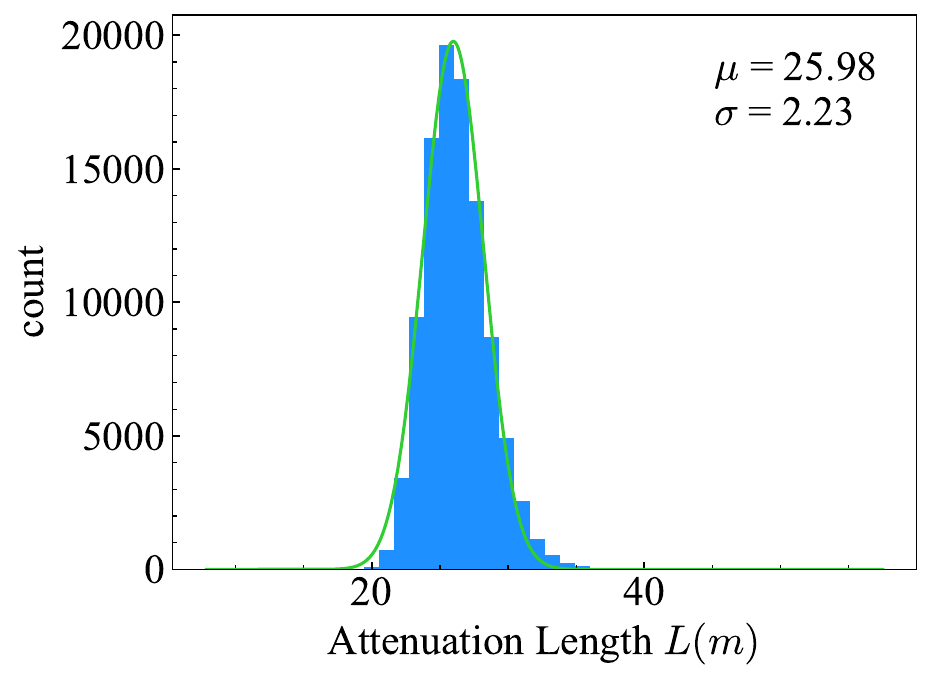}
        \caption{NJ44 before upgrades}
        \label{fig:MC NJ44 before}
    \end{subfigure}
    \begin{subfigure}{0.45\textwidth}
        \includegraphics[width = \textwidth]{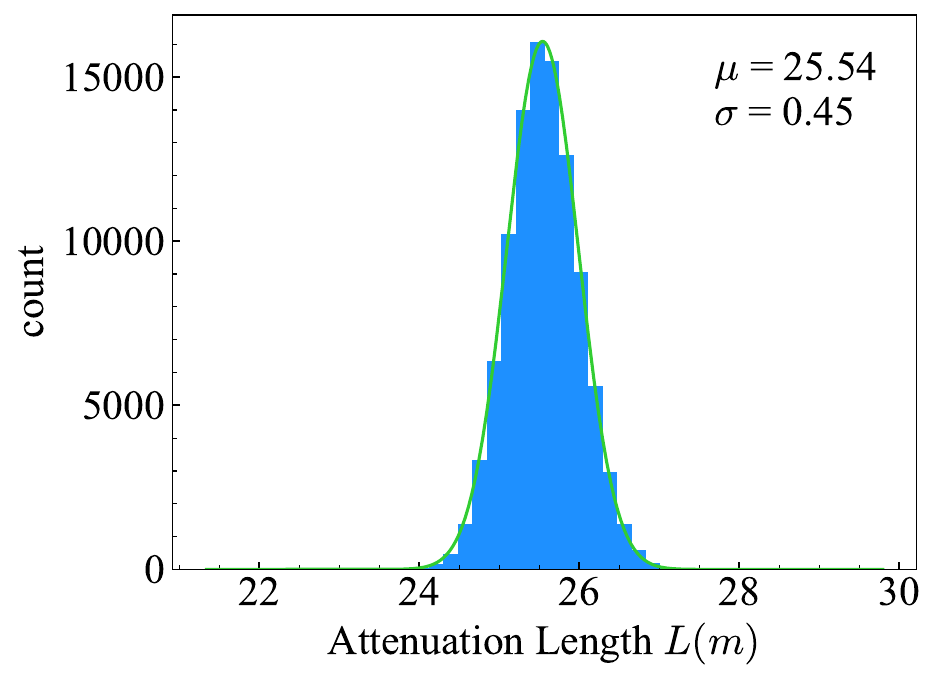}
        \caption{NJ44 after upgrade}
        \label{fig:MC NJ44 after}
    \end{subfigure}
    \begin{subfigure}{0.45\textwidth}
        \includegraphics[width = \textwidth]{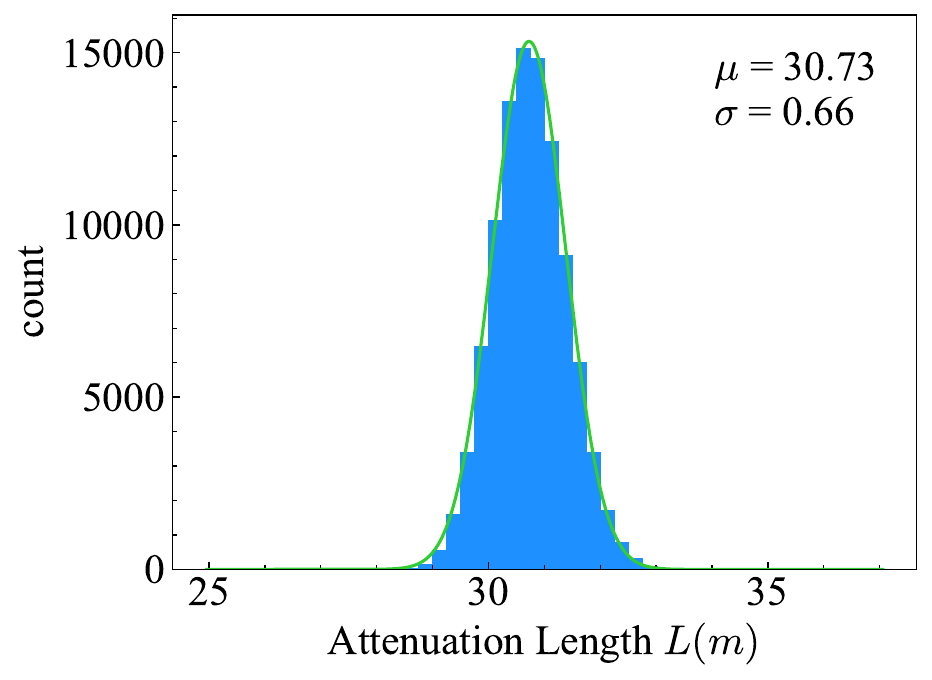}
        \caption{NJ66 test 1}
        \label{fig:MC NJ66 1}
    \end{subfigure}
    \begin{subfigure}{0.45\textwidth}
        \includegraphics[width = \textwidth]{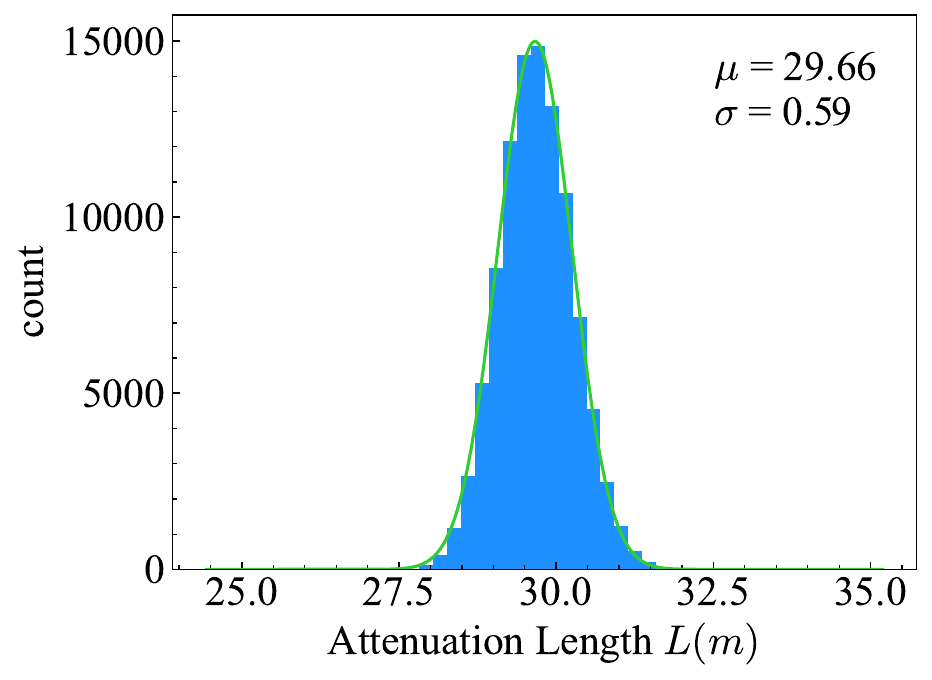}
        \caption{NJ66 test 2}
        \label{fig:MC NJ66 2}
    \end{subfigure}
\caption{Distributions of \(L\) simulated by Monte Carlo method. The upper two plots presents a comparison of systematic uncertainties of NJ44 before and after the upgrades. The lower two are simulations run for two independent tests of the same sample NJ66.}
\label{fig:MC}
\end{figure*}

\begin{figure*}[htp!]
\centering
    \hspace{3mm}
    \begin{subfigure}{0.425\textwidth}
        \includegraphics[width = \textwidth]{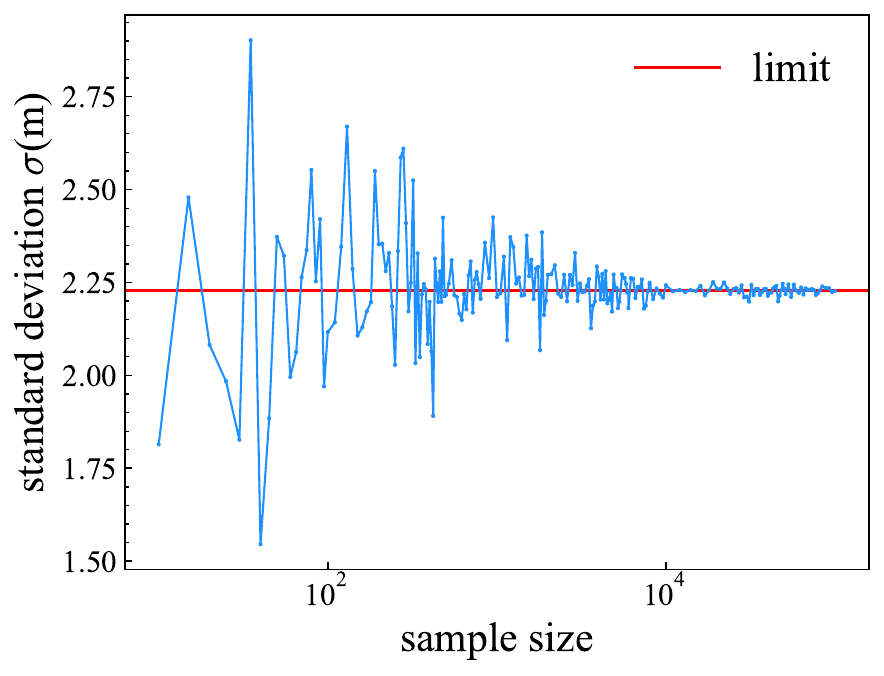}
        \caption{NJ44 before upgrades}
        \label{fig:MC convergence before}
    \end{subfigure}\hspace{6mm}
    \begin{subfigure}{0.42\textwidth}
        \includegraphics[width = \textwidth]{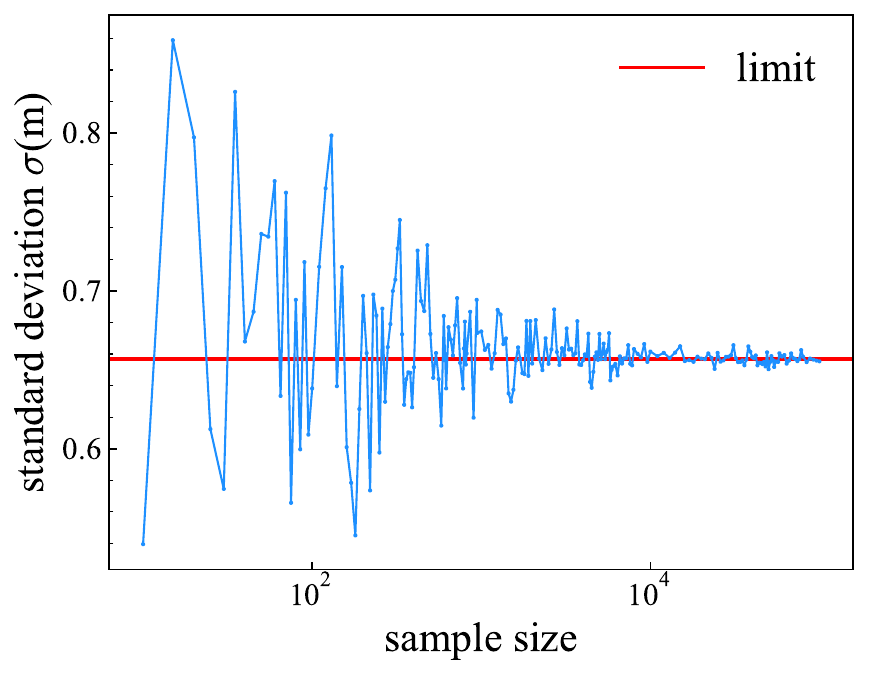}
        \caption{NJ66 after upgrades}
        \label{fig:MC convergence after}
    \end{subfigure}
\caption{The standard deviation of simulated \(L\) converges when the number of \(L\) is large enough. Convergence limit has been set to be the average of the last 50 values.}
\label{fig:MC convergence}
\end{figure*}

Now that \(\varepsilon_{I_i}\) and \(\varepsilon_{x_i}\) are known, \(\varepsilon_L^{\text{sys}}\) can be determined. Figure \ref{fig:MC} shows results of Monte Carlo simulations with a sample size of \(100,000\). Figure \ref{fig:MC NJ44 before} and \ref{fig:MC NJ44 after} are distributions of NJ44 MC simulated data before and after upgrades. A reduce in the standard deviation after the upgrade is impossible to ignore. Figure \ref{fig:MC NJ66 1} and \ref{fig:MC NJ66 2} are distributions of two measurements of NJ66. A convergence test has also been conducted for NJ66 after upgrades, showing the limit of \(\varepsilon_L^{\text{sys}}\) in figure \ref{fig:MC convergence}. Here we claim the systematic uncertainty of each measurement of NJ66 to be \(0.66\metre\) and \(0.59\metre\) respectively. The overall uncertainty of both systematic and statistical uncertainty is shown in figure \ref{fig:results} so the statistical uncertainty can be calculated through the square sum relation of errors. We claim the statistical uncertainty of each measurement of NJ66 to be \(1.28\metre\) and \(0.63\metre\) respectively.

\subsection{Errors of modeling}

In light of complex interactions of photons and sample substance, Beer--Lambert law is only valid when some strict conditions are met. Any imperfection of apparatus would result deviations from the ideal exponential law. This inevitable deviation becomes more significant when other errors get suppressed through the apparatus upgrade. In this experiment, the fact that residuals show a common pattern and reduced chi-squares of fit is higher than usuals clearly demonstrates the existence of higher order terms, implying a deviation from the fit model. As is mentioned earlier the absorption in LAB is mainly caused by impurities and it is Rayleigh scattering that dominates the attenuation process,\cite{Huang2011, Xiao2010} we propose a modified model giving special concern on the Rayleigh scattering.

\begin{figure}[htp!]
    \centering
    \includegraphics[scale = 0.15]{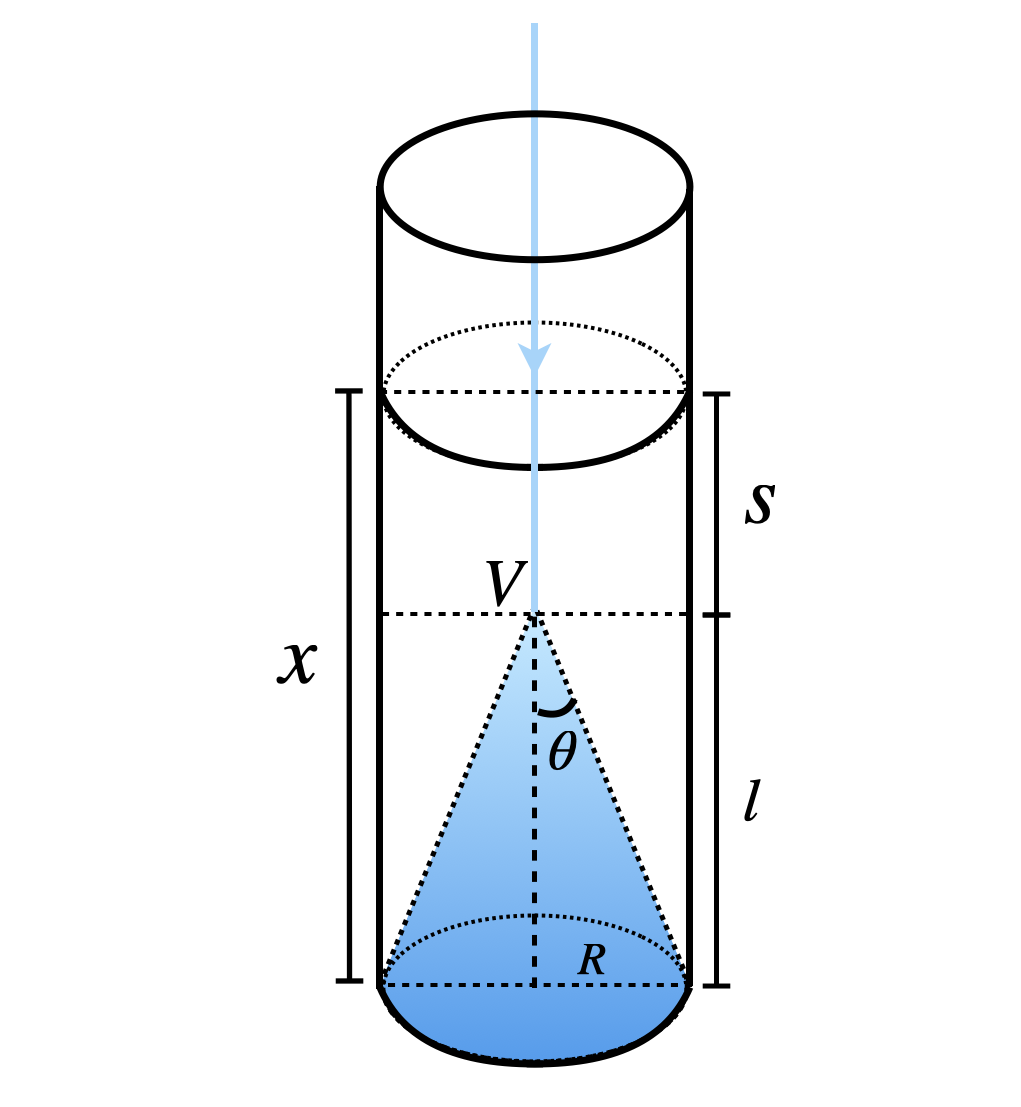}
    \caption{Geometry of light passing through sample in the one-metre-tube. The blue cone shows the solid angle in which the scattering light can be received.}
    \label{fig:geometry}
\end{figure}

Rayleigh scattering, as is first proposed in 1871 by Lord Rayleigh,\cite{rayleigh1871-1, rayleigh1871-2, rayleigh1881, rayleigh1889} is an elastic process, which means it only changes photons' direction without changing its frequency. Thus it is natural to assume that the high on both ends shown in residual plots originates from the downward part of scattering light that can be received by the PMT. To testify this hypotheses let us trace back to the grounding assumption of deriving Beer--Lambert law that for a thin layer the ratio of the attenuated to the incident is a constant \(\mu_{\text{att}}\), which is exactly the inverse of attenuation length \(\mu_{\text{att}} = 1/L\). We correct this by simply adding a term \(\mu_{\text{mod}}\) representing the not lost part of the scattering light,

\begin{equation} 
    \frac{1}{I}\frac{\md I}{\md s} = -\mu_{\text{att}} + \mu_{\text{mod}}
\end{equation}
where \(I\) is intensity and \(s\) is the length light already traveled in sample. \(\mu_{\text{mod}}\) is determined by the geometry of the tube shown in figure \ref{fig:geometry}. For tubes in a cylindrical shape as used in this experiment, only the part of scattering light whose scattering angle is within the solid angle subtended by the tube's bottom end window can be received. Thus, the not lost part takes a proportion of
\begin{equation} 
    \eta(s) = \int_0^{\alpha(s)}\md\theta\int_0^{2\pi}\md\varphi\,f(\lambda, \theta, r)\,r^2\sin\theta
\end{equation}
in which \(\varphi,\,\theta,\, r\) is local spherical coordinate, \(l\) the length from scattering vertex to the bottom end of the tube and \(s = x - l\) with \(x\) being the liquid level, \(\alpha(s) = \arctan(R/l)\) is the corresponding maximal azimuthal angle. \(R\) the radius of the bottom window. Integrand \(f(\lambda, \theta, r)\) is the Rayleigh formula for unpolarized incident beams\cite{Stewart1925}
\begin{equation} 
    f(\lambda, \theta, r) = \frac{2\pi^2(n-1)^2}{N\lambda^4r^2}(1+\cos^2\theta)
\end{equation}
where \(n\) is the refractive index, \(N\) the number density of the sample, \(r\) the distance from the scattering vertex and \(\lambda\) the wavelength of the incident beam. Recently some researches shows that organic liquid such as LAB is not isotropic and depolarized part also contributes to the scattering.\cite{Liu2015, Zhou2015-1, Zhou2015-2,Yu2022} Thus, a depolarized parameter \(\rho\) should be introduced. J. Cabannes and L.V. King generalized Lord Rayleigh's formula to the anisotropic molecules of gases and liquids,\cite{Cabannes1920, King1923} where 
\begin{equation} 
    f(\lambda, \theta, r) = \frac{\pi^2(n-1)^2}{2N\lambda^4r^2}\cdot\frac{6+6\rho}{6-7\rho}\left(1+\frac{1-\rho}{1+\rho}\cos^2\theta\right)
\end{equation}
Here \(b = \pi^2(n-1)^2/(2N\lambda^4)\) is introduced to function as the new fit parameter and \(\rho = 0.31 \pm 0.02\) is adopted in the new model.\cite{Liu2015}

This part of scattering light attenuates as well going through rest of the sample, an damping term is inevitable.
\begin{equation} 
    \mu_{\text{mod}}(s) = \me^{-\mu_{\text{att}}(x-s)}\eta(s)
\end{equation}
To simplify the tedious integration the exponential is firstly expanded and then cut off. Since higher order terms in the expansion stands for the strongly attenuated part it has little impact on fit and the first two terms would suffice to show the physics. The corrected model is finally obtained and denoted as

\begin{equation} 
    I(x) = I_0\me^{-\mu_{\text{att}} x + \int_0^x\mu_{\text{mod}}(s)\,\md s}
\end{equation}

\begin{figure*}[htp!]
\centering
    \begin{subfigure}{0.47\textwidth}
        \includegraphics[width = \textwidth]{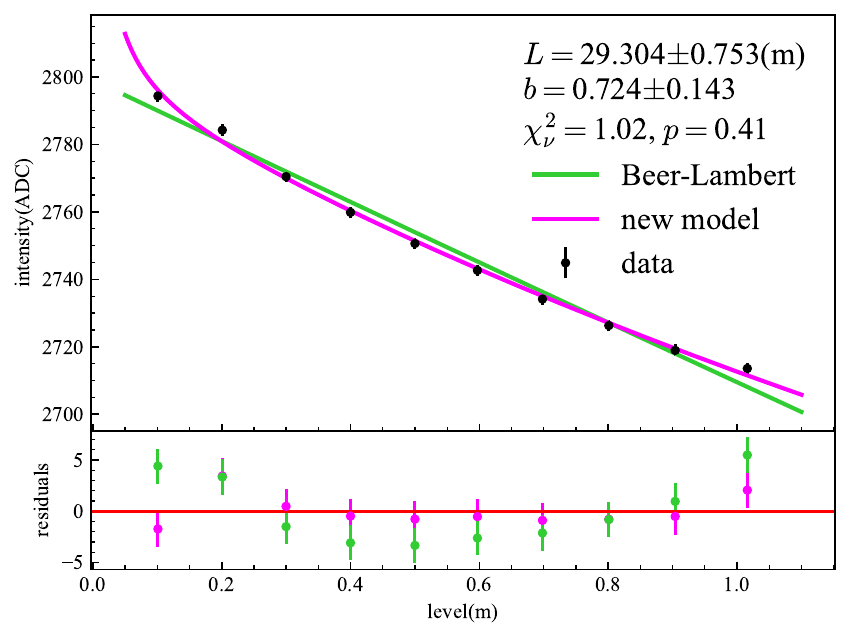}
        \caption{NJ66 test 1}
        \label{}
    \end{subfigure}
    \begin{subfigure}{0.47\textwidth}
        \includegraphics[width = \textwidth]{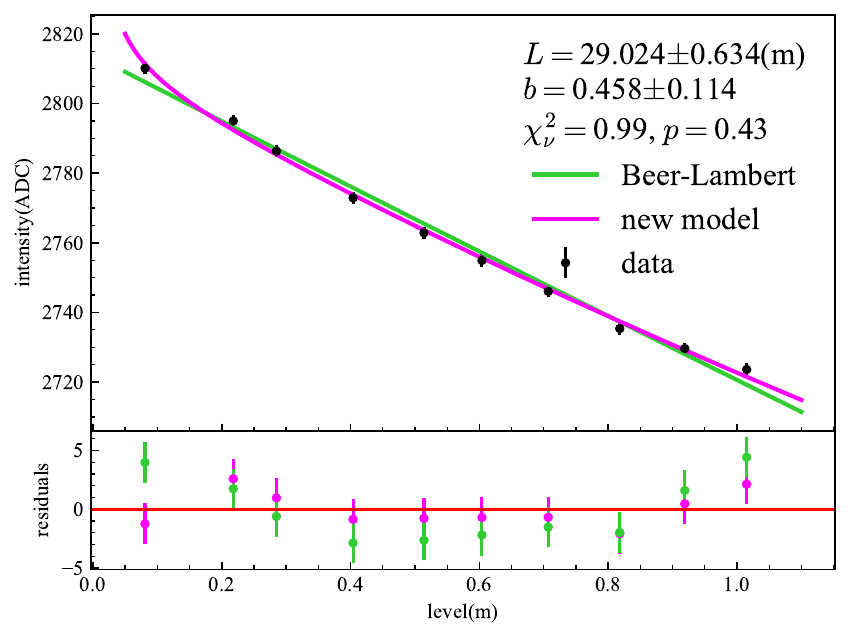}
        \caption{NJ66 test 2}
        \label{}
    \end{subfigure}
\caption{Fits using the new model. Reduced chi-square \(\chi_{\nu}^2\) has seen a decrease to close to 1 and most of the data points are located to the model expectations within 1 error bar.}
\label{fig:modeling}
\end{figure*}

This new model is fit to the same datasets of samples like the old model. Here results of two independent tests of NJ66 are presented in figure \ref{fig:modeling}. As is drawn from the plots, the \(\varepsilon_L^{\text{model}}\) for these two measurements are respectively \(1.43\metre\) and \(0.65\metre\). Quantitatively, \(\chi^2_{\nu}\) decreasing to near \(1\) and \(p\) value close to \(0.5\) imply a better goodness-of-fit of the new model than that of Beer--Lambert law which works only in an ideal situation. However, it has to be pointed out that for residuals there still seems to be a pattern when \(x > 0.4\metre\) and the seemingly soaring behavior near \(x \rightarrow 0\metre\) awaits for deeper scrutiny as well. This may be compensated by involving more terms when cutting off the exponential but may also indicate some other factors not taken into account in this physical analysis such as contributions of diffuse reflection and integrand's dependence on the wavelength.

\subsection{Combining}

Since we have conducted two independent measurements of NJ66 sample, a combined result considering both measurements can be given. The two independent results are presented as follows:
\begin{equation} 
\begin{split}
    &L_1 = 30.73\pm 0.66^{\text{sys}}\pm 1.28^{\text{sta}}\pm 1.43^{\text{model}}\metre\\% = 30.73\pm 2.03\metre \\
    &L_2 = 29.66\pm 0.59^{\text{sys}}\pm 0.63^{\text{sta}}\pm 0.65^{\text{model}}\metre\\% = 29.66\pm 1.08\metre
\end{split}\end{equation}

And the the combined result is given as the weighted sum

\begin{equation} 
    L = \frac{L_1/\varepsilon_1^2 + L_2/\varepsilon_2^2}{1/\varepsilon_1^2 + 1/\varepsilon_2^2} = 29.90 \metre
\end{equation}
with an uncertainty of

\begin{equation} 
    \varepsilon_L = \frac{1}{\sqrt{1/\varepsilon_1^2 + 1/\varepsilon_2^2}} = 0.95 \metre
\end{equation}

Finally, the light attenuation length of sample NJ66 is reported to be \(29.90 \pm 0.95 \metre\).

\section{Conclusions}

In this article the attenuation length of 4 batches of LAB samples has been measured and analysed in details. It is explained why an upgrade of apparatus is inevitable at present. The collimated fibre coupled light source that shrinks the size of spot together with a PMT with better time performance increases measuring limit of attenuation length and decreases systematic error of intensity to a large extent. Fit data show that after the upgrade the apparatus has received a substantial improvement in both accuracy and precision. We also proposed an elaborated uncertainty approximation method featuring Monte Carlo simulation to estimate systematic uncertainty, saving us from complicated calculations. By proposing a new model, deviations of theoretical models caused by Rayleigh scattering has also been considered. We eventually report a result of an attenuation length of the most updated LAB sample NJ66 to be \(29.90 \pm 0.95 \metre\). We hope that not only these results help to give more confidence on JUNO reaching its energy resolution but also the statistical method and the theoretical model exhibited in this article could be inspiring for similar experiments.

\begin{acknowledgments}
    This work was supported by the National 973 Project Foundation of the Ministry of Science and Technology of China (Contract No.2013CB834300), the Strategic Pilot Science and Technology Project of the CAS (Contract No.XDA10010000), and the National Natural Science Foundation of China (Contract No.11620101004).
\end{acknowledgments}

\section*{Data Availability Statement}
The data that support the findings of this study are available from the corresponding author upon reasonable request.

\bibliography{reference}

\end{document}